\documentclass{mem}
\usepackage{natbib}
\usepackage{balance}
\usepackage{txfonts}
\usepackage{graphicx}
\usepackage[a4paper]{hyperref}
\idline{75}{282}
\begin{document}

\title{The VO-Neural project:}

\subtitle{recent developments and some applications}

\author{M. \,Brescia\inst{1}, S. \,Cavuoti\inst{2}, G. \,D'Angelo\inst{2}, R. \,D'Abrusco\inst{2}, N. \,Deniskina\inst{2},\\ 
M. \,Garofalo\inst{3},  O. \,Laurino\inst{2}, G. \,Longo\inst{2}$^,$\inst{1}, A. \,Nocella\inst{3}, B. \,Skordovski\inst{2}}

  \offprints{M. Brescia}

\institute{
INAF - Osservatorio Astronomico di Capodimonte, via Moiariello 16, 80131, Napoli,
\email{brescia@na.astro.it}
\and 
Department of Physical Sciences - University Federico II, Naples, Italy
\and
Department of Computer Engineering - University Federico II, Naples, Italy}

\authorrunning{M. Brescia et al. }

\titlerunning{VO-Neural project:}

\abstract{VO-Neural is the natural evolution of the Astroneural project which was started in 1994 
with the aim to implement a suite of neural tools for data mining in astronomical massive data sets. 
At a difference with its ancestor, which was implemented under Matlab, VO-Neural is
written in C++, object oriented, and it is specifically tailored to work in distributed computing architectures. 
We discuss the current status of implementation of VO-Neural, present an application to the
classification of Active Galactic Nuclei, and outline the ongoing work to improve
the functionalities of the package.

}

\maketitle{}

\section{Introduction}

One of the main goals of the International Virtual Observatory (VOb) is the federation under common standards 
of all astronomical archives available worldwide \cite{IVOA}. 
Once this meta-archive will be completed, its exploitation will allow a new type of multi-wavelenght, multi-epoch 
science which can only be barely imagined \cite{george_1}, but will also pose unprecedented computing problems. 
From a mathematical point of view, in fact, most of the operations performed by the astronomers 
during their every-day life can be reconduced (either consciously or unconsciously) to standard data 
mining tasks such as, for instance, clustering, classification, pattern recognition and trend analysis. 
All these tasks scale very badly when either the number $N$ of records to be processed or the number $D$ of 
features characterizing each record, increase: 
\begin{itemize}
\item clustering scales as $\sim N \times \log N \times N^2$, and as $\sim D^2$;
\item search for correlations scales as $\sim N \times \log N \times N^2$, and as $\sim D^k$ with  $k \geq 1$;
\item bayesian or likelihood  algorithms scale as $\sim N^m$ with $m\geq 3$ and as$\sim D^k$ with  $k \geq 1$.
\end{itemize}
To get an idea of the computational demands posed by the VOb we shall just notice that a modern digital survey can 
easily produce datasets having $N\sim 10^{9}$ and $D\gg 10^2$ and leave to the reader to imagine what could be the 
demands of a multiwavelenght, multi-epoch survey.
It is apparent that the extraction of knowledge from such data sets cannot be performed with traditional SW\cite{astroweka} \& HW, and 
requires some form of high performance computing (HPC). 
The traditional HPC approach based on parallel multi-CPU software running on dedicated clusters, is however against the 
very same phylosophy of the VOb which aims at opening the exploitation of its data archives also to scientists who do not 
have access to large HPC centers. 
In this respect, the GRID seems to offer the most natural and democratic answer since, at least in theory, it allows any 
user possessing a personal certificate to access the distributed computing resources. 
The VOb, however, for the same fact of being open to use by the community at large, does not match the security 
requirements of the GRID and this limitation strongly undermines its effectiveness.

In \cite{deniskina2008} we discuss the first version of $GRID-Launcher$, a tool which interfaces the UK-ASTROGRID \cite{astrogrid} 
with the GRID-SCOPE \cite{scope}. 
In this contribution we discuss instead the structure of the data mining package VO-Neural \cite{voneural} 
which is specifically designed to perform complex data mining (DM) tasks on astronomical (but not
only) massive data sets (MDS).
As an exemplification, in Sect.\ref{applications} we also show how the methods so far implemented can be used to address the 
challenging task of obtaining an objective classification of Active Galactic Nuclei (AGN).
Finally, in the last Section we shortly outline some ongoing and planned developments.

\section{VO-Neural}
VO-Neural is a data mining framework, whose goal is to provide the astronomical community with powerful software instruments 
capable to work on massive ($>1$ TB) data sets (catalogues) in a distributed computing environment matching the IVOA standards and 
requirements. 
VO-Neural is the evolution of the AstroNeural \cite{astroneural} project which was started in 1994, as a collaboration 
between the Department of Mathematics and Applications at the University of Salerno and the Astronomical Observatory of Capodimonte-INAF, 
and is currently under continuous evolution.
VO-Neural allows to extract from large datasets information useful to determine patterns, relationships, similarities and regularities 
in the space of parameters, and to identify outlayers. 
In its final version, it will offer main elaborative features like exploratory data analysis, data prediction and ancillary functionality 
like fine tuning, visual exploration of the main characteristics of the datasets, etc..
Besides offering the possibility to use the individual routines to perform specific tasks, VO-Neural will
provide the user with a complete framework to write his own customized programs.

\noindent Without entering into too many details we shall just recall that, in our view, data exploration means agglomerative clustering 
and dimensional reduction of parametric space; data prediction means prediction, classification and regression; 
fine tuning means Not a Number (NaN) or upper limits determination and outlayers, catalogue statistical analysis and data extraction. 

\begin{figure*}[t!]
\resizebox{\hsize}{!}{\includegraphics[clip=true,width=10cm]{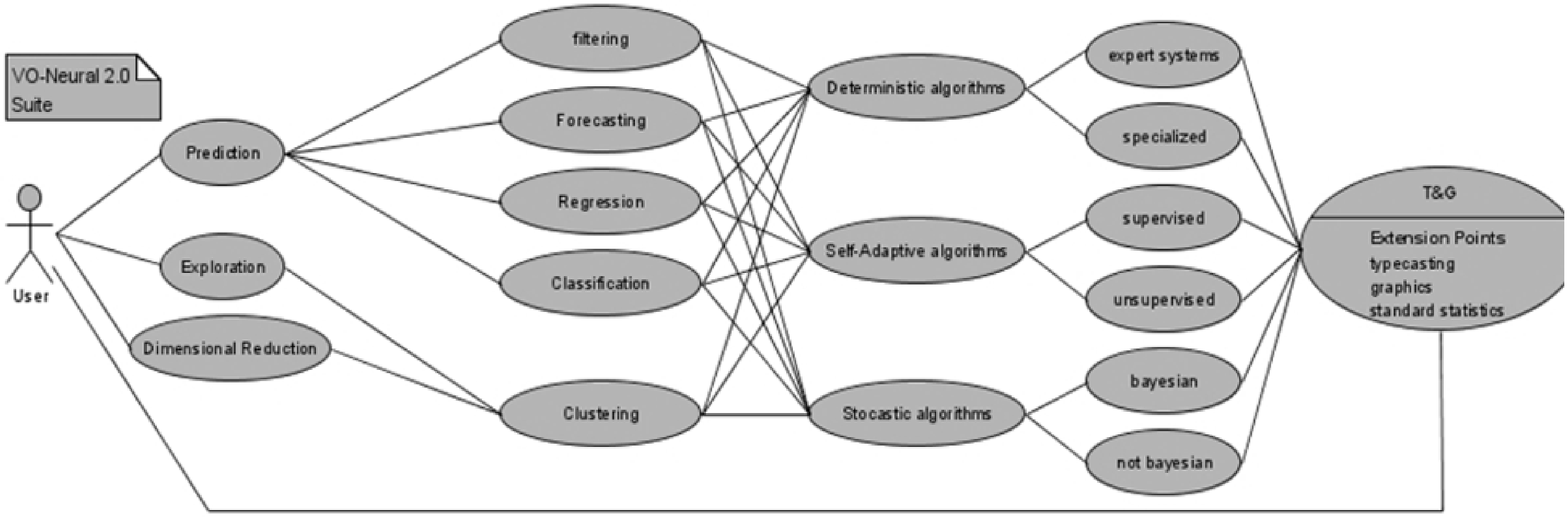}a)} 
\resizebox{\hsize}{!}{\includegraphics[clip=true,width=10cm]{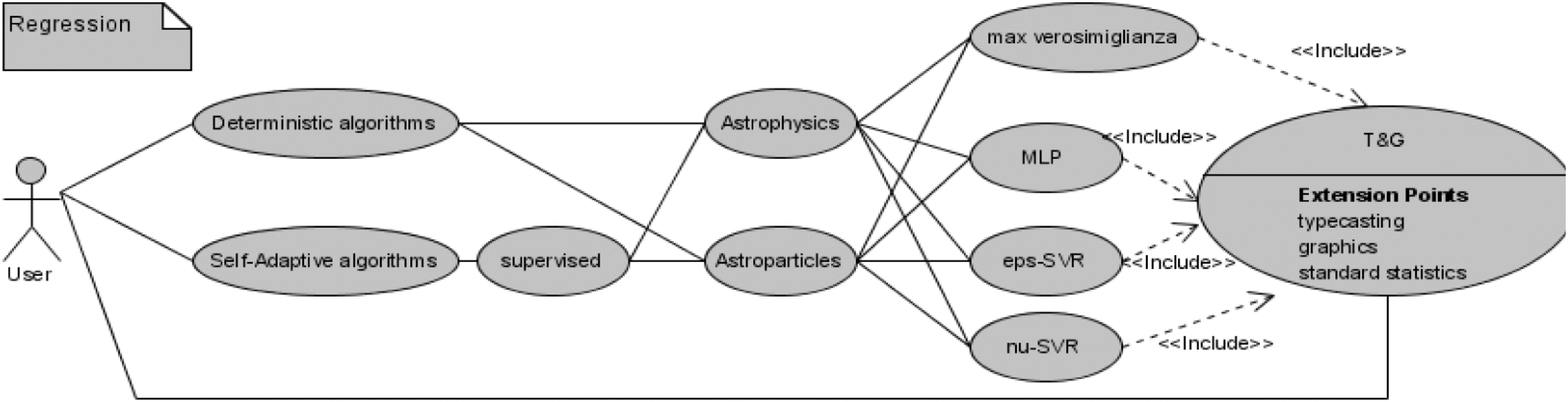}b)} 
\resizebox{\hsize}{!}{\includegraphics[clip=true,width=10cm]{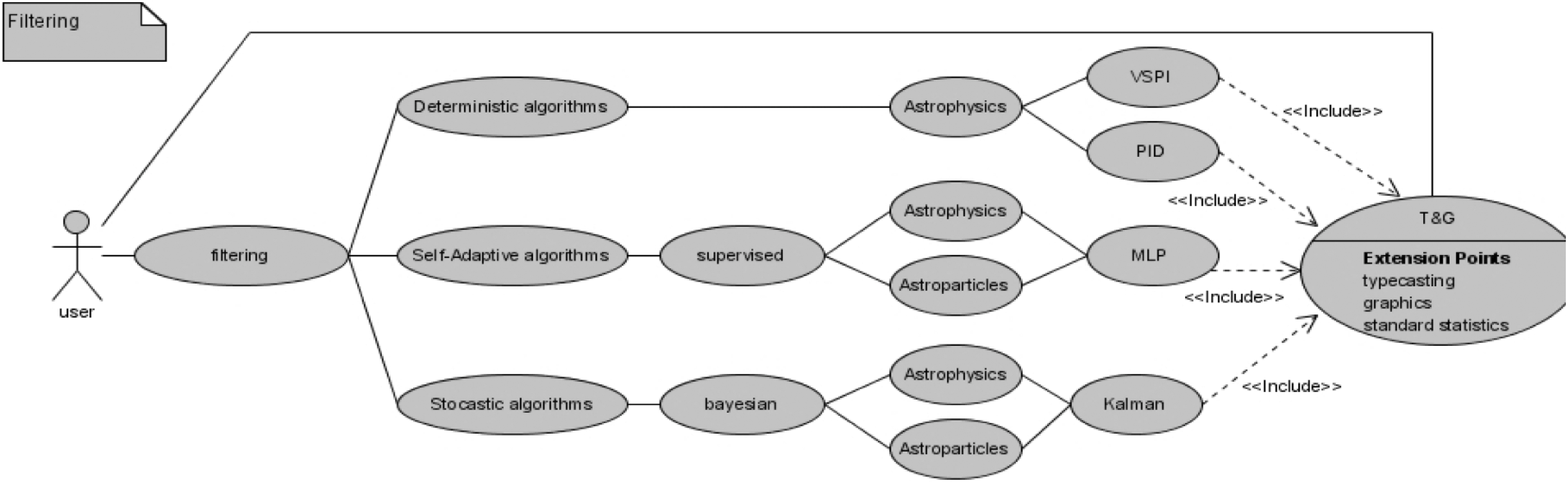}c)} 
\resizebox{\hsize}{!}{\includegraphics[clip=true,width=10cm]{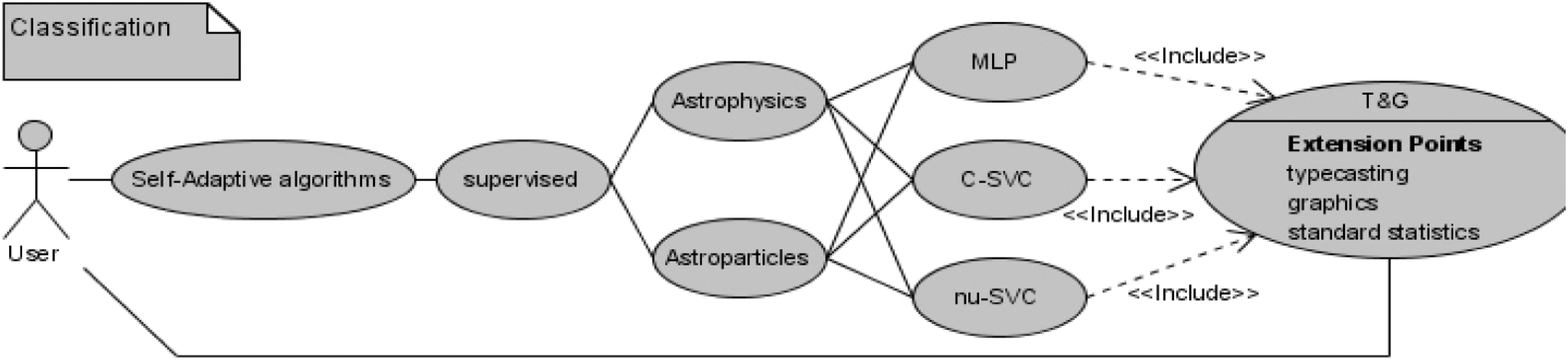}d)} 
\caption{\footnotesize a): logical flow of the VO-Neural package; b-c-d) explosion of some subsections
of the package.} \label{figcf}
\end{figure*}

With reference to Fig.\ref{figcf} we specify that deterministic, self-adaptive and statistical methods are 
implemented to achieve the above functions requirements as embedded in a generic pipeline.
Deterministic models include triggers and data reduction algorithms. 
Self-adaptive models are organized in supervised and unsupervised tools. 
Statistical models refer to simple statistical functions, either Bayesian or not-Bayesian and 
dimensional reduction models to clustering methods like Probabilistic Principal Surfaces (PPS) and Negative Entropy Clustering (NEC).
Classification includes self-adaptive models like supervised neural network (MLP with back-propagation and genetic 
algorithms, C-SVC and NU-SVC) and, finally, regression refers to Multi Layer Perceptron, other supervised self-adaptive models, 
like EPSILON-SVR and NU-SVR, and to data fitting deterministic algorithms.
Moreover a set of graphical analysis tools (such as histograms and wisker \& bar plot, etc.) is included.

VO-Neural is built around the following standards:
\begin{itemize}
\item XP-agile as suite designing method;
\item UML (Unified Modeling Language);
\item OOP (Object Oriented Programming);
\item interface protocols based on EGEE, VO \& AstroGrid paradigms;
\item standard I/O interface methods for software systems integrity;
\item SVN (SubVersion) software version for control \& archiving;
\item webservice-based user interfaces.
\end{itemize}

In the next two paragraphs we shortly outline the main features of two supervised clustering
models already included in the package which have already been used for specific science applications.

\subsection{$VONeural\_MLP$}
$VONeural\_MLP$ is an implementation of a standard Multi Layer Perceptron based on the FANN 
(Fast Artificial Neural Networks) Library \cite{fann}, written in C \cite{skordovski},
and tailored to be launched as web service from the ASTROGRID Workbench.
The algorithm known as Multi Layer Perceptron (MLP) is based on the concept of perceptron
and the method of learning is based on gradient-descent method that allows to
find a local minimum of a function in a space with N dimensions. 
The weights associated to the connections between the layers of neurons, initialized at small and random 
values, and then the MLP applies the learning rule using the template patterns.
\subsection{$VONeural\_SVM$}
$VONeural\_SVM$ is an implementation of the Support Vector Machines \cite{russo,cavuoti} based on
the LIBSVM library \cite{libsvm}. 
Support Vector Machines perform classification of records into classes by first mapping the data into an higher dimensionality
and then using a set of template vectors (targets) to find in this new space an iperplane of separation with the largest margin possible.
Withouth enetering into details (which can be found in (Boser et al., 1992; Cortes and Vapnik, 1995), we shall just remember that, 
in the case of the C-SVC implemented with the RBF (Radial Basis Functions) kernel, the position of this hyperplane depends on two 
parameters ($C$ and $\Gamma$) which cannot be estimated in advance but need to be evaluated by finding the maximum in a grid of 
values which is usually defined by letting $C$ and $\Gamma$ vary as $C = 2^{-5}, 2^{-3}, ..., 2^{15}$ and $\Gamma = 2^{-15}, 2^{-13}, ..., 2^3$.
Due to their computational weight, and to the need to run many iterations for different pairs of the two parameters,
SVM are ideally suited for the GRID.

\section{The classification of AGN}\label{applications}

The astronomical community is used to perform DM tasks in a sort of ''hidden'' way (cf. the case of specific objects 
selection in a color-color diagram) but it has not yet become familiar with the potentialities of more advanced tools 
such as those described here.
This is mainly due to the fact that these tools are often everything but user friendly and require an in depth 
understanding of the (often complex) theory laying behind them; a complexity which often discourages potential users.
Therefore, a crucial aspect of the project is the application to challenging problems which, 
can better exemplify the new science which will emerge from the adoption of a less 
conservative approach to the analysis of the data.
Two science cases, namely the evaluation of photometric redshifts (a regression and classification problem
based on the use of $VONeural\_MLP$) and the selection of candidate quasars in the Sloan Digital Sky Survey \cite{SDSS}
(based on the use of unsupervised clustering algorithms and agglomerative clustering) have already been published 
in the literature \cite{dabrusco,dabrusco_1,dabrusco_2}. 
We shall therefore focus on a new application of $VONeural\_SVM$ to the classification of AGNs.

\noindent The classification of AGN is usually performed on their overall spectral distribution using some spectroscopic 
indicators (equivalent linewidths, FWHM of specific lines or lines flux ratios) and diagnostic diagrams (usually called BPT).
In this diagrams AGN and not-AGN are empirically separated by some lines derived either from the theory or from
empirical laws such as those derived by \cite{kewley,kauffman,heckman}.
A reliable and accurate AGN classificator based on photometric features only, would allow to save precious telescope time 
and enable several studies based on statistically significant samples of objects. 
We therefore used a supervised clustering approach based on a Base of Knowledge (BoK) derived from the available catalogues. 
We wish to stress that since neural networks have no power of extrapolation all the biases in the BoK will be reproduced in 
the final results.
As classification tools, we used the MLP and, due to the intrinsically binary nature of the problem (AGN against non-AGN, 
Seyfert 1 against Seyfert 2, etc) also the SVM. 
\begin{table*}\label{results}
\begin{tabular}{llcrr}
\hline
\hline
experiment       & BoK                    & algorithm & efficiency & completeness\\
\hline
AGN vs Mix       & BPT plot + Kewley line & MLP       & $76\%$     & $54\%$      \\
                 & BPT plot + Kewley line & SVM       & $74\%$     & $55\%$      \\
\hline
Type 1 vs 2      & BPT plot + Kewley line & MLP       & $95\%$     & $\sim 100\%$\\
                 & BPT plot + Kewley line & SVM       & $82\%$     & $98\%$      \\
\hline
Seyfert vs LINER & BPT plot + Hecman \& Kewley lines & MLP       & $80\%$     & $92\%$      \\
                 & BPT plot + Kewley line & SVM       & $78\%$     & $89\%$      \\
\hline
\end{tabular}
\caption{Summary of the results of supervised classification experiments performed using both 
$VONeural\_MLP$ and $VONeural\_SVM$.}
\end{table*}
\noindent The BoK was obtained from the fusion of two catalogues.
\begin{itemize}
\item \cite{sorrentino} separated objects into Seyfert 1, Seyfert 2 and ''Not AGN'' 
using the Kewley's lines \cite{kewley}; 
\item a catalogue derived by us from the SDSS spectroscopic archive using the criteria introduced by \cite{kauffman} 
in which objects are classified as AGN, not AGN, and ''mixed''. 
The Mix and Pure AGN zone were further divided into Seyfert and LINERs by using the Heckman line\cite{heckman}.
\end{itemize}
\noindent We made three experiments using both the MLP and SVM, and for all of them we used the same set of features 
(for a definition refer to the SDSS specifications) extracted from the SDSS database: 
$petroR50\_u$, $petroR50\_g$, $petroR50\_r$, $petroR50\_i$, $petroR50\_z$, $concentration\_index\_r$, $fibermag\_r$, 
$(u-g) dered$, $(g-r) dered$, $(r-i) dered$, $(i-z) dered$, $dered\_r$, together with the photometric redshift 
in \cite{dabrusco_1}.
We performed three types of classification experiments: AGN vs Mix, Type1 vs Type2, Seyfert vs LINER. 
The experiments with SVM were performed on the GRID-SCOPE using 110 worker nodes. 
The results are summarized in Table \ref{results}.

As it can be seen, the use of machine learning tools allows to reach performances which 
in some cases (e.g. Type 1 vs 2 with MLP's) cannot by any means be achieved with more traditional tools. 
A more detailed discussion of the results will be presented in (Cavuoti, d'Abrusco \& Longo, 2008, in preparation).

\section{Future developments}
The ongoing work is focused on three main aspects: i) implementing better methods 
through an extensive parallelization of the already existing codes; ii) improving the
interfacement of the package with the GRID; 
iii) incorporating within the VO-Neural package tools capable to extract information 
from the data collected from the new generation of astroparticle physics experiments.
\\

\noindent {\it Acknowledgements.} The authors wish to thank M. Paolillo and E. de Filippis for many
useful discussions. The work was funded through the Euro VO-Tech project and the MUR funded PON-SCOPE. 

\bibliographystyle{aa}

\end{document}